# The Casimir Atomic Pendulum

**H. Razmi** [(1)] and **M. Abdollahi** [(2)]

Department of Physics, The University of Qom, Qom, I. R. Iran.
(1) Email: razmi@qom.ac.ir & razmiha@hotmail.com
(2) Email: mah.abdollahi@gmail.com

## Abstract

We want to introduce an atomic pendulum whose driving force (torque) is due to the quantum vacuum fluctuations. Applying the well-known Casimir-Polder effect to a special configuration (a combined structure of an atomic nanostring and a conducting plate), an atomic pendulum (Casimir atomic pendulum) is designed. Using practically acceptable data corresponding to the already known world of nanotechnology and based on reasonable/reliable numerical estimates, the period of oscillation for the pendulum is computed. This pendulum can be considered as both a new Micro (Nano)-Eelectromechanical system and a new simple vacuum machine. Its design may be considered as a first step towards realizing the visualized vacuum (Casimir) clock!

**Introduction**

Is it possible to extract energy and/or exploit from the infinite resource of quantum vacuum "sea"? Although this is a controversial subject among present experts, some ones believe in it and try to "make" vacuum machine [1]. The Casimir effect [2] is known as a real manifestation of the "abstract" concept of quantum vacuum. Most of researches corresponding to this effect deal with the Casimir energy and/or the Casimir force for different geometrical configurations under particular conditions [3]; there are a few publications directly deal with the visualized vacuum machine (e.g. papers dealing with the subject of the Casimir torque [4-5], an interesting but controversial paper introducing vacuum battery [6], and another work dealing with the anharmonic Casimir oscillator [7]). J. Schwinger believed in this idea that the sonoluminoscence phenomenon [8] originates from the Casimir effect for a dielectric ball [9]; of course, there are some serious criticisms on this idea (K. Milton and others [3, 10-11] believe this phenomenon is irrelevant to the Casimir effect because if Schwinger had considered the bulk energy contribution in his formula, his result would have been much smaller than what needed to justify sonoluminescence). In 1996, some members of NASA defined a project named as BPP (Breakthrough Propulsion Physics) with the aim of designing a rocket whose driving force is to be extracted from quantum vacuum!; they haven't found any hopeful result yet [12-13].

We are going to introduce/design a pendulum that works based on the Casimir effect. The special configuration we consider here is completely different from the "Casimir pendulum" introduced in [14]. The authors of [14] have studied the interaction of two plates using both an *ad hoc* PFA (proximity force approximation) method and an optical approach. Here, using the Casimir-Polder effect for a moving atom, we want to study the oscillation of an atomic nanostring based upon a conducting plate. We use practically possible data estimates both in the design of the pendulum and in the numerical computation of its period of oscillation. This may be considered as a first step towards having/making a quantum vacuum (Casimir) clock!

## The atom-plate interaction

Consider a polarized atom near a conducting plate; because of the applied boundary condition (the confinement of spatial TE and TM electromagnetic modes), there is an attractive force due to the interaction of atom with quantum vacuum electromagnetic field. This is the so-called Casimir-Polder effect [15]. Frequently, this effect is studied based on the assumption that the atom is static; if the atom is let to go (adiabatically) away from the conducting plate, the static Casimir-Polder effect needs some corrections [16]. Indeed, in this case, there is a restoring force that gets back the atom towards the plate. In reference [16], it has been shown that if the atom is released infinitely far from the conducting plate and moves in toward it, then the restoring force near the plate will be twice the stationary value; this is the maximum value. Using this fact that if the atom is released but remains stationary, then the restoring force will be equal to the stationary atom value, the restroring force can be considered as ( $F_R \approx \beta F_{CP}$, $1 \leq \beta \leq 2$ ). The asymptotic form of the potential influences the static atom is:

$$U(R) \to -\frac{\alpha_0 \hbar \omega_0}{32\pi} \frac{1}{R^3}, \qquad R \ll \frac{c}{\omega_0} \qquad (1)$$

$$U(R) \to -\frac{3\alpha_0 \hbar c}{32\pi^2} \frac{1}{R^4}, \qquad R \gg \frac{c}{\omega_0} \qquad (2)$$

where $\omega_0$ is the transition frequency of the atom, $R$ is the distance from the atom to the plate, and $\alpha_0$ is the static polarizability of the atom.

## The pendulum configuration

Consider a (mono)atomic nanostring of length $l$ (a linear chain of about 30 atoms) whose last atom is polarized. It is hung at a distance $d$ above a conducting plate (see figure 1). Initially, the string is inclined from the vertical line to a very small angular value $\varphi$. Using this condition and because of the small (below-mentioned) numerical value of $R$, we can use the near zone approximation (1). In this approximation, as is clear from (1), the $R$ distance is much smaller than the characteristic transition wavelength of the atom and thus the potential is essentially of electrostatic origin. Because of its small length, the nanostring can be assumed as a rigid string [17]. The forces acting on the system are the Casimir-Polder force $(F_{CP})$, the restoring force $(F_R)$, and the gravitational force acting on the center of mass of the string $(W = Mg)$:

$$F_{CP} = -\frac{\partial}{\partial R}U(R) = -\frac{3\alpha_0 \hbar \omega_0}{32\pi}\frac{1}{R^4}, \qquad R << \frac{c}{\omega_0} \qquad (3)$$

$$F_R \approx \beta F_{CP}, \quad 1 \leq \beta \leq 2 \qquad (4)$$

Now, let estimate the real possible numerical values of the problem. The typical value of transition frequency of an atom is of the order of $\omega_0 \approx 10^{15} s^{-1}$; so, $(R << \frac{c}{\omega_0})$, the distance $R$ is considered $R << 3\times 10^{-7} m$. The number of atoms the nanostring made of them is considered about 30 ones; this is in agreement with already known values [17]. If one assumes the center to center of any two neighboring atoms is about three times the atom radius (see figure 2), then:

$$r \approx 10^{-10} m \rightarrow l = 30 \times 3 \times 10^{-10} \approx 10^{-8} m \qquad (5)$$

Therefore, the nanostring mass is estimated as:

$$m = \frac{atomic\ weight}{6.022 \times 10^{23}} \approx 10^{-25} kg \rightarrow M = 30 \times 10^{-25} \approx 10^{-24} kg \qquad (6)$$

The atomic polarizability whose value is of the order of atom "volume" is estimated as ($\alpha \approx 10^{-30}$).

It should be mentioned that if one wants to design such a configuration in the far zone limit, because of both the greater value of $R$ and the different functional form of the Casimir-Polder force ($\propto 1/R^5$), the estimated numerical value of the forces under consideration will be very small and even negligible (Note: we would like the main driving force/torque of the pendulum to be due to the Casimir effect).

**The pendulum period**

The equation of motion for the atomic pendulum is

$$Mg \frac{l}{2} \sin\varphi + \frac{3(1+\beta)\hbar\omega_0 \alpha}{32\pi(d-l\cos\varphi)^4} l \sin\varphi + I\ddot{\varphi} = 0 \quad , \quad 1 \leq \beta \leq 2 \quad (7)$$

where $I \cong Ml^2/3$ is the moment of inertia.

A comparison of two first terms of the above equation based on the real numerical estimates of the previous section, assuming $R \approx 10^{-8} m$, shows that the torque due to gravitational effect is negligible:

$$Mg \frac{l}{2} \approx 10^{-31} \, N/m, \quad \frac{3(1+\beta)\hbar\omega_0 \alpha l}{32\pi(d-l\cos\varphi)^4} \approx 10^{-26} \, N/m \quad (8)$$

Thus:

$$\frac{3(1+\beta)\hbar\omega_0 \alpha}{32\pi(d-l\cos\varphi)^4} l \sin\varphi + \frac{Ml^2}{3} \ddot{\varphi} = 0 \quad (9)$$

For reasonable small values of the angle of oscillation $\varphi$ ($\sin\varphi \cong \varphi, \cos\varphi \cong 1$):

$$\ddot{\varphi} + \frac{9(1+\beta)\hbar\omega_0 \alpha}{32\pi Ml(d-l)^4} \varphi = 0 \quad (10)$$

Therefore, the angular velocity and period of oscillation are found as:

$$\omega = \sqrt{\frac{9(1+\beta)\hbar\omega_0 \alpha}{32\pi Ml(d-l)^4}} \to T = 2\pi \sqrt{\frac{32\pi Ml(d-l)^4}{9(1+\beta)\hbar\omega_0 \alpha}} \quad (11)$$

Using the above estimated data, one arrives at the following result:

$$T \approx 10^{-7} s \quad (12)$$

This period is of the order of one tenth of microsecond which can be considerable in the micro/nanoworld. Since the driving force (torque) of the pendulum is due to the quantum vacuum fluctuations, we have named this system as "the Casimir atomic pendulum". It can be considered as both a new Micro (Nano)-Eelectromechanical system and a new simple vacuum machine. It may be also considered as a first step towards realizing the visualized vacuum (Casimir) clock! One may ask about the possibility of the experimental verification; what we can say is that because of the simplicity of the design and practically acceptable data corresponding to the already known world of nanotechnology (in particular about the rigidity of the nanostring), it seems we aren't so far away from the starting point of having a suitable set-up for being practically experienced.

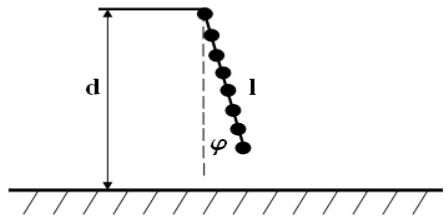

*Figure 1: The pendulum configuration*

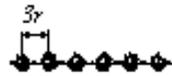

*Figure 2: A linear atomic nanostring*